\pdfoutput=1
\documentclass [11pt,letterpaper]{JHEP3}
\usepackage{amsmath}
\usepackage{bm}         
\usepackage{verbatim}   
\usepackage{graphicx}
\advance\parskip 1.0pt plus 1.0pt minus 2.0pt   
\def\coeff#1#2{{\textstyle {\frac {#1}{#2}}}}

\def\half{\coeff 12}
\def\l{\ell}
\def\C{{\cal C}}
\def\N{{\cal N}}

\def\S_1{{\widetilde {S_1}}}

\def\R{{\mathbb R}}

\def\tr{{\rm tr}}

\def\x{\mathbf x}
\def\r{\mathbf r}

\def\Nc{N_{\rm c}}
\def\Lc{L_{\rm c}}
\def\Nc{N}
\def\Nf{N_{\rm f}}
\def\Z{{\mathbb Z}}

\def\Dslash{{\rlap{\raise 1pt \hbox{$\>/$}}D}}
\def\O{{\cal O}}

\def\betatilde{\widetilde\beta}

\def\Im{\mathrm {Im}}

\preprint{SLAC-PUB-13144\\NSF-KITP-08-16}

\title
    {%
    \boldmath
    Center-stabilized Yang-Mills theory:
    confinement and large $N$ volume independence
    }%

\author
    {
    {
    \def\href#1#2{#2}	
    Mithat \"Unsal%
    $^1$\footnote{\email{unsal@slac.stanford.edu}}~
    and Laurence G.~Yaffe%
    $^2$\footnote{\email{yaffe@phys.washington.edu}}
    \\${}^1$SLAC and Physics Department, Stanford University, Stanford, CA 94305
    \\${}^2$Department of Physics, University of Washington,
    Seattle, Washington 98195--1560
    }
    }%

\abstract
    {%
    We examine a double trace deformation of $SU(N)$ Yang-Mills theory
    which, for large $N$ and large volume, is equivalent to unmodified
    Yang-Mills theory up to $O(1/N^2)$ corrections.
    In contrast to the unmodified theory,
    large $N$ volume independence is valid in the deformed theory
    down to arbitrarily small volumes.
    The double trace deformation prevents the spontaneous breaking
    of center symmetry which would otherwise disrupt large $N$ volume
    independence in small volumes.
    For small values of $N$, 
    if the theory is formulated on $\R^3 \times S^1$
    with a sufficiently small compactification size $L$,
    then an analytic treatment of the non-perturbative dynamics of
    the deformed theory is possible.
    In this regime, we show that the deformed Yang-Mills theory
    has a mass gap and exhibits linear confinement.
    Increasing the circumference $L$ or number of colors $N$ decreases
    the separation of scales
    on which the analytic treatment relies.
    However, there are no order parameters which distinguish the small
    and large radius regimes.
    Consequently,
    for small $N$ the deformed theory provides a novel example
    of a locally four-dimensional pure gauge theory
    in which one has analytic control over confinement,
    while for large $N$ it provides a simple fully
    reduced model for Yang-Mills theory.
    The construction is easily generalized to QCD
    and other QCD-like theories.
    }%

\begin{document}

\section {Large $\bm N$ volume independence}

The large $N$ limit of $SU(N)$ Yang-Mills theories,
when formulated on toroidal compactifications of $\R^d$,
are independent of volume provided the $(\Z_N)^d$ center symmetry
is not spontaneously broken
\cite{Eguchi-Kawai,LGY-largeN,BHN}.%
\footnote
    {
    Center symmetry transformations are gauge transformations
    which are periodic only up to an element of the center of
    the gauge group.
    Volume independence applies to the leading large $N$
    behavior of expectation values and connected correlators
    of topologically trivial Wilson loops.
    }
However, above two dimensions,
center symmetry {\em does} break spontaneously
when the (smallest) compactification circumference $L$ is less than
a critical size $\Lc$ \cite{Narayanan-Neuberger}.
(If just one dimension is compactified, then this
center-symmetry breaking transition is the usual
thermally induced deconfinement transition.)
In four dimensions, 
the critical size $\Lc$ is approximately $\Lambda^{-1}$
where $\Lambda$ is the $\overline {MS}$ strong scale of the theory
\cite{Narayanan-Neuberger, Kiskis-Narayanan-Neuberger}.

Notwithstanding the limitation to $L \ge \Lc$,
the volume independence of large $N$ Yang-Mills theory
(``partial reduction'')
has practical utility for lattice studies,
because simulations on lattices of size $(\Lc)^d$ are sufficient to extract
infinite volume properties of large $N$ Yang-Mills theory
\cite{Narayanan-Neuberger, Lucini:2005vg, Bringoltz:2005xx, Cohen:2004cd}.
But it would be even more helpful to have a formulation of
the theory in which volume independence holds for arbitrarily small volumes ---
since this allows one to reduce the lattice all the way down to a single site.

Several schemes for preserving volume independence in arbitrarily
small volumes have been proposed.
In so-called quenched reduced models, one constrains the eigenvalues of
link variables (or Wilson lines) in a manner which prevents Wilson lines
from acquiring expectation values \cite{BHN}.%
\footnote
    {
    See Ref.~\cite{Makeenko} for an extended discussion of quenched
    and twisted reduced models.
    }
In the $N \to \infty$ limit, quenched reduced models correctly reproduce
properties of infinite volume Yang-Mills theory.
However, corrections to the $N = \infty$ limit scale as
$1/N$, not $1/N^2$, in quenched reduced models.
This makes extracting large $N$ properties from numerical simulations
of quenched reduced models quite challenging.
An alternative proposal, known as twisted reduced models,
involves modifying the Wilson action of a single-site model
so that the action explicitly disfavors configurations in which
Wilson lines in different directions mutually commute
\cite{Gonzalez-Arroyo:1982hz, Gonzalez-Arroyo:1982ub}.
Unfortunately, this clever scheme fails to work sufficiently
close to the continuum limit \cite{Teper:2006sp, Azeyanagi:2007su}.
In essence, the penalty imposed by the twisting of the action
is insufficient to overcome entropic effects which favor
breaking of the center symmetry.

If light adjoint representation fermions are added to an $SU(N)$
Yang-Mills theory,
and periodic (not anti-periodic) boundary conditions imposed on the fermions,
then the fermion contribution to the Wilson line
effective potential stabilizes the unbroken center symmetry phase.
Hence these QCD-like theories satisfy large-$N$ volume independence
for arbitrarily small volumes \cite{Kovtun:2007py}.
(In addition, in the large-$N$ limit
$\C$-even observables coincide between these theories and
corresponding theories, in sufficiently large volume,
with fermions in the rank-two symmetric or antisymmetric
tensor representations \cite{UY,ASV1, Armoni:2004ub}.)

Motivated by this fermion-induced stabilization of center symmetry,
in this paper we introduce a simple scheme for preserving
volume independence in pure Yang-Mills theory.
We add double trace terms to the action
which prevent spontaneous breaking of center symmetry,
while simultaneously perturbing the dynamics of the
unbroken symmetry phase only by $O(1/N^2)$ corrections.
This leads to a ``stabilized reduced model'' which
reproduces the dynamics of infinite volume Yang-Mills
theory up to corrections which scale as $1/N^2$.
The construction may be easily generalized
to other QCD-like theories with matter fields in
rank-one or rank-two representations. 

In addition to providing a simple large $N$ reduced model,
the deformed Yang-Mills theory is interesting in its own right
when $N$ is not large.%
\footnote
    {
    See also related recent work in 
    Refs.~\cite{Schaden:2004ah, Pisarski:2006hz, Myers:2007vc, Shifman:2008ja}.
    }
When formulated on $\R^3 \times S^1$,
we show that the large distance dynamics of the theory
is analytically tractable provided $N \Lambda L \ll~1$.
In this regime, a semiclassical analysis (closely related to Polyakov's
classic treatment of $3d$ $SU(2)$ adjoint Higgs theory \cite{Polyakov:1976fu})
reveals the existence of a mass gap and area law behavior of
spatial Wilson loops.
It is noteworthy that our compactified, deformed Yang-Mills theory
is an analytically tractable confining theory 
with no fundamental scalar fields or supersymmetry,
in contrast to other instructive models of confinement
\cite{Polyakov:1976fu,Seiberg:1994rs,Douglas:1995nw,Deligne:1999qp}.
The confinement mechanism involves the formation of a dilute plasma of
magnetic monopoles (and antimonopoles) carrying topological charge $\pm 1/N$.%
\footnote
    {
    Confinement due to such topological objects has been previously discussed
    in, for example, Refs.~\cite{Zhitnitsky:2006sr,Toublan:2005tn}
    and references therein.
    What is novel about our deformed theory at small $N\Lambda L$
    is that this confinement mechanism operates in a regime in which one has
    analytic control over the long distance dynamics.
    }

\section {Deformed Yang-Mills theory}

We consider pure Yang-Mills (YM) theory with gauge group $SU(N)$ defined on
the four manifold  $\R^3 \times S^1$,
with the $S^1$ having circumference $L$.
The extension to multiple compactified
dimensions will be discussed below, but we begin with a single
compactified dimension to simplify the exposition.
We start with the usual continuum action,
\begin{equation}
    S^{\rm YM}= \int_{\R^3 \times S^1} \frac{1}{2g^2} \>
    \tr\,  F_{\mu \nu}^2 (x) \,,
\label{eq:cont}
\end{equation}
or a lattice formulation with the Wilson action,
\begin{equation}
    S^{\rm YM}
    = \frac\beta 2 \sum_{p\in \Lambda_4}
    \tr\, \bigl( U[\partial p] + U[\partial p]^\dagger \bigr) ,
\label{eq:lattice}
\end{equation}
where the sum is over all oriented plaquettes,
$\Lambda_4$ is the four dimensional spacetime lattice,
and $U[\partial p]$ denotes the usual product of link
matrices around the boundary of plaquette~$p$.
The lattice coupling $\beta \equiv 2/g^2$.
In our discussion, we will use both continuum and lattice formulations,
and benefit from both perspectives.
As usual, a key virtue of the lattice formulation is that it provides
an explicit non-perturbative definition of the theory.

Let
$\Omega({\bf x}) \equiv \mathcal P \> (e^{i\int dx_4 \> A_4({\bf x}, x_4)})$
denote the Wilson line (or Polyakov loop) operator ---
the holonomy of the gauge field around a circle wrapping
the $S^1$ and sitting at the point ${\bf x} \in \R^3$.
We will construct a deformation of the Yang-Mills action
on our compactified geometry by adding terms, respecting all symmetries
of the unmodified theory,
built from the Wilson line operator.
The deformed action is given by
\begin{equation}
    S^{\rm deformed} = S^{\rm YM} + \Delta S \,,
\end{equation}
with
\begin{subequations}
\label{eq:Delta S}
\begin{align}
    \Delta S
    &\equiv
    \int_{\R^3}  \frac{1}{L^3} \> P[\Omega({\bf x})]
\\
\noalign{\hbox{in the continuum, or}}
    \Delta S
    &\equiv
    \frac{1}{N_t^3} \, \sum_{{\bf x} \in \Lambda_3}  P[\Omega({\bf x}) ]
\end{align}
\end{subequations}
on the lattice.
In the lattice form,
$N_t \equiv L/a$ denotes the size of the lattice in the compactified direction
and $\Lambda_3 \subset \Lambda_4$ is a three dimensional
sublattice of the four dimensional lattice
corresponding to a fixed Euclidean time-slice.
We want the deformation potential $P[\Omega]$
to guarantee the stability of the phase with unbroken center symmetry
(at small volume).
It will be chosen to have the form
\begin{equation}
    P[\Omega] \equiv \sum_{n=1}^{\lfloor {N/2} \rfloor}
    a_n \left|\tr \left( \Omega^n \right) \right|^2 \,,
\label{eq:P[Omega]}
\end{equation}
with positive coefficients $\{ a_n \}$
(and $\lfloor N/2 \rfloor$ denoting the integer part of $N/2$).
In other words, $P[\Omega]$ is
a sum of the double trace operators
$\tr (\Omega^n) \, \tr (\Omega^n)^\dagger$.
When considering the large $N$ limit, the coefficients $\{ a_n \}$
will be held fixed as $N \to \infty$.

Under a center symmetry transformation 
by some element $z \in \Z_N$, 
the Wilson loop $\tr(\Omega^p)$
is multiplied by $z^p$.
The value of $p \bmod N$, which determines the $\Z_N$ representation,
is referred to as the $N$-ality.
It will be important that
the deformation potential (\ref{eq:P[Omega]}) only contains,
by construction,
absolute squares of Wilson loops with non-zero $N$-ality.

If $P[\Omega]$ were simply proportional to $|\tr \, \Omega|^2$,
with a sufficiently large positive coefficient,
then this would prevent breaking of the center symmetry with
$\langle \tr (\Omega) \rangle$ as an order parameter.
But if $N > 3$ then this single term is not sufficient to prevent
any spontaneous breaking of center symmetry,
as this term alone does nothing to prevent
$\tr (\Omega^2)$ from developing an expectation value.
In other words, a stabilizing term proportional to $|\tr (\Omega)|^2$
cannot prevent $\Z_N$ breaking to $\Z_2$ (assuming $N$ is even)
with $\tr(\Omega^2)$ as an order parameter.
Adding an additional stabilizing term proportional to $|\tr(\Omega^2)|^2$
could prevent such a breaking to $\Z_2$, but 
does not prevent breaking to $\Z_3$ (if $N \bmod 3 = 0$),
or to any larger discrete subgroup of $\Z_N$.
This is why we have allowed $P[\Omega]$ to include terms up to
$|\tr\, \Omega^{\lfloor N/2 \rfloor}|^2$.

We will argue that the deformed theory satisfies the following:
\begin{itemize}
\item[{\it i})]
    For suitable choices of the deformation parameters $a_n$
    ({\em i.e.}, each coefficient sufficiently large and positive)
    the stabilizing potential (\ref{eq:P[Omega]})
    will prevent the $\Z_N$ center symmetry from breaking to any subgroup.

\item[{\it ii})]
    In the $N\to\infty$ limit, pure Yang-Mills theory on $\R^4$
    is equivalent to the deformed theory formulated on any $\R^3 \times S^1$
    (for choices of the $\{a_n\}$ satisfying point {\em i}).
    This equivalence applies to expectation values of Wilson loops
    (on $\R^4$), or the leading large $N$ behavior of their
    connected correlators.
    In the lattice formulation, the number of sites in the
    compactified direction may be reduced to one.

\item[{\it iii})]
    When $N \Lambda L\ll1$ limit, the deformed Yang-Mills theory
    is solvable in the same sense as the Polyakov model.
    The existence of a mass gap and linear confinement
    can be shown analytically. 
    One can regard this regime as having spontaneous breaking of the
    $SU(N)$ gauge symmetry down to $U(1)^{N-1}$, but this is a perturbative
    gauge dependent description with no well-defined invariant content.

\item[{\it iv})]
    There exist no order parameters which can distinguish the
    $N\Lambda L\ll 1$ ``Higgs'' regime from the $N \Lambda L\gg 1$
    regime in which gauge symmetry is ``restored''.
\end{itemize}

\begin{FIGURE}[ht]
    {
    \parbox[c]{\textwidth}
        {
        \begin{center}
        \includegraphics[width=0.9\textwidth]{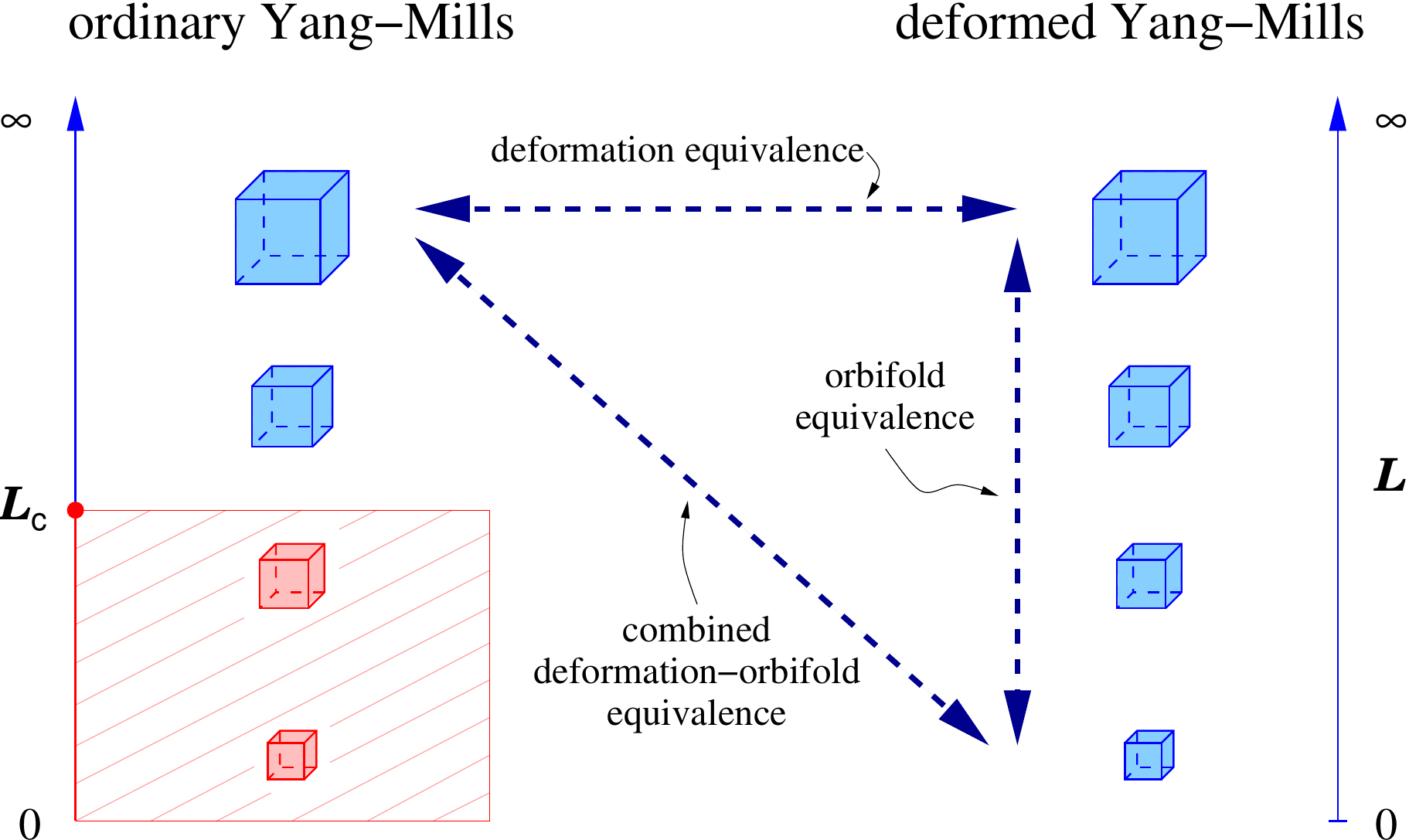}
        \caption
	{
	Large $N$ equivalences relating ordinary and deformed $SU(N)$
	Yang-Mills theories,
	as a function of the size $L$ of the periodic volume
	in which the theories are defined.
	In deformed YM volume independence holds for all $L$,
	while in ordinary YM volume independence fails below a critical
	size, $L<\Lc$, (shaded region)
	due to spontaneous breaking of center symmetry.
	This prevents reduction all the way down to a single-site matrix model
	for ordinary YM.
	Large $N$ equivalence holds between ordinary and deformed YM
	theories as long as center symmetry is unbroken in ordinary YM.
	Large $N$ volume independence is a type of orbifold equivalence,
	as discussed in Ref.~\cite{Kovtun:2007py}.
	The combination of volume changing orbifold projections in 
	the deformed theory,
	along with the deformation equivalence in sufficiently large volume,
	provides a useful equivalence between deformed YM in small volume
	and ordinary YM in large volumes.
	In particular, a single-site matrix model of the deformed theory
	will reproduce properties of ordinary Yang-Mills theory in
	infinite volume.
	The construction can be generalized to QCD,
	with the deformed theory providing a
	fully reduced matrix model for QCD. 
	}
        \label {fig:mappings}
        \end{center}
        }
    }
\end{FIGURE}

As noted earlier, pure Yang-Mills theory on $\R^3 \times S^1$
satisfies volume independence (in the large $N$ limit)
so long as the $\Z_N$ center symmetry remains unbroken 
\cite{LGY-largeN, BHN, Narayanan-Neuberger}.
We will show that
this is equally true for the deformed theory;
the additional terms in the action do not affect the proof
that volume independence, at $N = \infty$, is an automatic
consequence of unbroken center (and translation) symmetry.
In the undeformed theory,
the unbroken center symmetry phase is the low temperature confined phase,
$L >\Lc$ with
$\Lc \sim \Lambda^{-1}$ the inverse deconfinement temperature.
But in the deformed Yang-Mills theory, the stability of the
unbroken center symmetry phase is enforced by hand for all values of $L$.

Large $N$ volume independence, and
the large $N$ equivalence between ordinary Yang-Mills theory on $\R^4$
and the deformed theory on $\R^3 \times S^1$,
may be demonstrated by comparing Dyson-Schwinger equations 
({\em i.e.}, Migdal-Makeenko loop equations) for expectation values and
correlators of Wilson loops,
or alternatively by comparing the large $N$ classical dynamics
that may be derived by using appropriate large $N$ coherent states
\cite{LGY-largeN,LGY-largeN2,LGY-largeN3,KUY2}.
Figure~\ref{fig:mappings} summarizes the relation between
the large $N$ limits of ordinary and deformed Yang-Mills theories.
As long as the $\Z_N$ center symmetry is not spontaneously broken,
the dynamics of the theories defined by $S^{\rm YM}$ and
$S^{\rm deformed}$ are indistinguishable
at leading order in the $1/N$ expansion.
In particular, glueball
spectra of the two theories can differ only by order $1/N^2$ effects.
Agreement up to $O(1/N^2)$ terms also applies to the
string tension which characterizes the area law behavior
of large Wilson loops.

The large $N$ equivalence between ordinary Yang-Mills theory 
and the deformed theory in large volume,
combined with the large $N$ volume independence of the deformed theory,
circumvents the problems with previous formulations of reduced
models for pure Yang-Mills theory.
Unlike the original Eguchi-Kawai model \cite{Eguchi-Kawai}, 
its twisted variant \cite{Gonzalez-Arroyo:1982hz, Gonzalez-Arroyo:1982ub},
and the partial reduction of
Refs.~\cite{Narayanan-Neuberger, Kiskis-Narayanan-Neuberger},  
the equivalence to deformed YM theory remains valid in the limit of zero
compactification radius, irrespective of the value of the
(bare) gauge coupling.
And for finite $N$,
corrections to the large $N$ limit scale as $1/N^2$,
not $1/N$ as in quenched reduced models \cite{BHN}.
Consequently, it should be possible to study the deformed theory,
for relatively modest values of $N$ and vanishingly small volume,
and obtain accurate results for properties of ordinary Yang-Mills theory,
in the large $N$ limit, on $\R^4$.
As we discuss below, it is also instructive to study the
deformed theory for small values of $N$.
In this regime, it will be seen to provide a novel example
of a confining theory, only involving an $SU(N)$ gauge field,
which is analytically soluble.

\subsection {Stabilization of center symmetry}

The possibility of preventing spontaneous breaking of center symmetry
through the addition of a deformation potential of the form (\ref{eq:P[Omega]})
is largely self-evident.
A positive coefficient $a_n$ suppresses configurations in which
$\tr (\Omega^n)$ is non-zero.
Although the pure-gauge dynamics of the undeformed theory,
in small volume, leads to an effective potential for the Wilson line
which is minimized when $\Omega$ is an element of $\Z_N$,
adding the deformation potential $P[\Omega]$ changes
the shape of the Wilson line effective potential.
For sufficiently large values of the coefficients $\{ a_n \}$,
the effective potential will be minimized by configurations
in which $\tr (\Omega^n) =~0$ for $1 \le n \le \lfloor N/2 \rfloor$.
This implies that $\tr (\Omega^n) = 0$ for any integer $n$
which is non-zero modulo $N$ because,
for $SU(N)$-valued matrices,
$\tr (\Omega^n)$ is not independent of lower order traces
when $n > \lfloor N/2 \rfloor$.
Vanishing of these traces implies that
the eigenvalues of $\Omega$ are uniformly spaced
around the unit circle, so that the set of eigenvalues is
invariant under $\Z_N$ transformations (which multiply every eigenvalue
by $e^{2\pi i/N}$).
This shows that the center symmetry is not spontaneously broken.
Henceforth, we assume that the coefficients $\{ a_n \}$ of the
deformation potential $P[\Omega]$
{\em are\/} suitably chosen so as to enforce unbroken center symmetry
for all compactification radii.

This argument may be made much more explicit if
one considers small compactifications,
$L \ll \Lambda^{-1}$,
so that (due to asymptotic freedom)
the gauge coupling at the scale of the compactification is small 
and the theory is amenable to a perturbative treatment.
Quantum fluctuations generate a nontrivial potential for
the Wilson line \cite{Gross:1980br}.
In ordinary Yang-Mills theory,
integrating out the gauge field (and Faddeev-Popov ghosts)
produces the functional determinants
\begin{equation}
    [{\det}_{+} (-D^2_{\rm adj} \, \delta_{\mu \nu})]^{-1/2}  \;
    [ {\det}_{+}  (-D^2_{\rm adj}) ] = [ {\det}_{+}  (-D^2_{\rm adj})]^{-1} \,,
\end{equation}
where $\det_{+}$ denotes a determinant in the space of periodic functions
with period $L$.
Therefore, the  effective potential for the Wilson line is
\begin{equation}
 V[\Omega]= L^{-1}  \; \ln \; {\det}_{+} (-D^2_{\rm adj}) \,.
\label{eq:cw}
\end{equation}
For constant (or slowly varying) configurations,
the evaluation of the  functional determinant is straightforward
and yields
\cite{Gross:1980br}
\begin{align}
    V[\Omega]&=
    \int_{\R^3} \frac 1{L^4} \> \mathcal V[\Omega(x)] \,,
\\
\noalign{\hbox{with}}
    \mathcal V[\Omega] &\equiv
    -
    \frac{2}{\pi^2}
    \sum_{n=1}^{\infty}\frac{1}{n^4} \> 
    \left|\tr\, \Omega^n \right|^2 \,.
\label{eq:potential}
\end{align}
For sufficiently small $L$, corrections to this one-loop result
are negligible.
The effective potential (\ref{eq:potential}) is minimized when
the Wilson line is an element of the center,
$\Omega = e^{2\pi i k/N}$, $k = 0, \cdots, N{-}1$,
demonstrating the spontaneous breaking of $\Z_N$ symmetry,
in ordinary Yang-Mills theory, for sufficiently small compactifications.

To force unbroken center symmetry in the deformed theory,
the deformation potential $P[\Omega]$ must overcome the
effect of the one-loop potential (\ref{eq:potential}).
A simple specific choice for the deformation coefficients which,
in the continuum limit and for sufficiently large $N$,
accomplishes this is
$
    a_n = 4/ (\pi^2 \, n^4)
$.
For this choice, the deformation potential $\int_{\R^3} L^{-4} \,P[\Omega]$
is minus twice
the one-loop Wilson line effective potential (\ref{eq:potential})
of the undeformed theory, so the net effect of the deformation
is to flip the sign of the effective potential for the Wilson line.
The resulting combined potential is minimized when
$\tr\, \Omega^n = 0$ for all $n$ which are non-zero modulo $N$,
indicating unbroken center symmetry.

\subsection{Multiple compactified dimensions}

Instead of compactifying just one dimension,
one may consider the gauge theory on $\R^{4-d} \times T^d$,
where $T^d$ is a $d>1$ dimensional torus.
For simplicity, we will discuss the case of a symmetric torus
having size $L$ in each dimension.
Two particularly interesting cases are $d=3$, where the $L \to 0$ limit
will reduce to $SU(N)$ matrix quantum mechanics,
and $d=4$, where the zero size limit will reduce to a matrix model.

With multiple compactified dimensions, the center symmetry is $(\Z_N)^d$.
The preceding discussion generalizes in a straightforward fashion,
but there is one significant change: with multiple compactified dimensions
there is a richer set of possible symmetry realizations in which the
center symmetry is partially broken.
One needs to craft the deformation potential in such a fashion that
all unwanted symmetry realizations are suppressed.

With multiple compactified directions, one may define distinct Wilson
line operators $\Omega_1$, $\Omega_2$, ..., $\Omega_d$
each of which wraps one elementary cycle of the torus.
The classical Yang-Mills action suppresses configurations in which
Wilson lines in different directions are non-commuting.
(In other words, configurations with vanishing field strength
must have Wilson lines which commute and are covariantly constant.)
A deformation potential with the form (\ref{eq:P[Omega]}),
involving just one Wilson line $\Omega_i$, can prevent
symmetry realizations in which the $\Z_N$ center symmetry associated
with direction $i$ spontaneously breaks in a manner which is independent
of the $\Z_N$ symmetries associated with other directions.
However, it is also possible for the center symmetry to break in a manner
which leaves a unbroken subgroup involving correlated
symmetry transformations in different directions.
With, for example, $d = 2$, suppose that the $(\Z_N)^2$ symmetry breaks
to the diagonal $\Z_N$ subgroup.
Such a symmetry realization does not allow expectations of
$\tr\,\Omega_1^k $ or $\tr\,\Omega_2^k $ to be non-vanishing
(for any $k$ which is non-zero modulo $N$).
But this realization does allow $\tr\, \Omega_1 \Omega_2^\dagger$
to have a non-zero expectation value.
To prevent such a symmetry realization, it may be necessary to add a
$|\tr\,\Omega_1\Omega_2^\dagger|^2$ term to the deformation potential.

To see that this concern is not moot, it is instructive to examine
the one-loop effective potential for Wilson lines when there are
multiple compactified directions.
Once again,
if the physical size of the $d$-torus is much smaller than the inverse strong
scale $\Lambda^{-1}$,  then ``Kaluza-Klein'' modes with momenta
at and above the compactification scale can be integrated out perturbatively.
This produces a one loop effective potential for mutually commuting
Wilson lines given by \cite{Luscher:1982ma, vanBaal:1988va}
\begin{align}
    \mathcal V[\Omega_1, \cdots, \Omega_d ] &\equiv
    -
    \frac{1}{\pi^2}
    \sum_{(n_1, \cdots, n_d) \in (\Z^d - {\bm 0}) }^{\infty}
    \frac
    {\left|\tr\left(\Omega_1^{n_1} \cdots \Omega_d^{n_d} \right)\right|^2}
    {\left(n_1^2 + \cdots + n_d^2\right)^2}
    \,.
\label{eq:potential2}
\end{align}
This effective potential has $N^d$ degenerate global
minima at which each Wilson line is an independent element of the center.
This demonstrates the spontaneous breaking of $(\Z_N)^d$ symmetry
(down to the identity) in ordinary Yang-Mills theory
on $\R^{4-d} \times (S^1)^d$ for sufficiently small compactifications.%
\footnote
    {\label{thermo}%
    More precisely, this demonstrates the spontaneous breaking of
    $(\Z_N)^d$ symmetry down to the identity when $d=1$ or 2.
    If $d > 2$ then there are fewer than two non-compactified directions.
    In this case, the center symmetry cannot
    break spontaneously for any finite value of $N$,
    due to the non-zero probability of fluctuations (or tunneling events)
    which effectively average over all degenerate minima of the
    Wilson line effective potential.
    However, the probability of such fluctuations vanish
    exponentially with increasing $N$.
    Consequently, in the $N\to\infty$ limit one can have spontaneous
    breaking of the $(\Z_N)^d$ center symmetry even when all directions
    are compactified.
    In order to have a large $N$ equivalence between the compactified
    theory and Yang-Mills theory on $\R^4$, one must prevent spontaneous
    breaking of the center symmetry in the large $N$ limit.
    Therefore, the more complicated form of the stabilizing potential
    discussed below is necessary for all $d > 1$.
    }
Adding a deformation to the Yang-Mills action which is the direct
generalization of the previously discussed deformation (\ref{eq:Delta S}),
namely
\begin{equation}
    \Delta S = \sum_i \int_{\R^{4-d}} \frac 1{L^{4-d}} \; P[\Omega_i] \,,
\label{eq:P2}
\end{equation}
with $P[\Omega]$ having the form (\ref {eq:P[Omega]}),
will deform the effective potential and can prevent this symmetry realization
in which the $(\Z_N)^d$ symmetry breaks down to nothing.

However, the fact that all terms in the effective potential
(\ref{eq:potential2}) have negative coefficients demonstrates that the
center symmetry {\em will\/} partially break spontaneously even in the
presence of stabilizing terms of the form (\ref {eq:P2}).
In particular,
even when all traces wrapping a single compactified direction vanish,
the potential (\ref{eq:potential2}) favors
configurations in which traces wrapping multiple cycles
(such as $\tr\,\Omega_1\Omega_2^\dagger$) are non-zero
relative to the center-symmetry preserving configuration
we are trying to stabilize
(for which $\tr\,\Omega_1^{n_1}\cdots\Omega_d^{n_d} = 0$
for all $n_i$ that are non-zero modulo $N$).

Therefore, to prevent any spontaneous breaking of the center symmetry
when there are multiple compactified directions one must allow the
deformation of the action to include absolute squares of
order parameters which wrap multiple cycles of the torus.
In other words we need to have
\begin{equation}
    \Delta S
    =
    \int_{\R^{4-d}} \> \frac 1{L^{4-d}} \>
    P[\Omega_1,\cdots,\Omega_d] \,,
\end{equation}
where the deformation potential has the form
\begin{equation}
    P[\Omega_1,\cdots,\Omega_d]
    =
    \sum_{n_1,\cdots,n_d=-\lfloor N/2 \rfloor}^{\lfloor N/2 \rfloor}
    a_{n_1\cdots n_d} \,
    \left|
	\tr\,\left(\Omega_1^{n_1} \cdots \Omega_d^{n_d}\right)
    \right|^2 \,,
\label{eq:multi-dir P}
\end{equation}
with coefficients $a_{n_1\cdots n_d}$ which are
sufficiently large and positive.

A simple specific choice for the coefficients which,
in the continuum limit and for sufficiently large $N$,
accomplishes this is
$
  a_{n_1\cdots n_d} = \frac 2{\pi^2} \left( n_1^2 + \cdots + n_d^2 \right)^{-2}
$.
For this choice, the deformation potential
is minus twice
the one-loop effective potential (\ref{eq:potential2})
of the undeformed theory, so once again the net effect of the deformation
is to flip the sign of the effective potential for the Wilson line.
The resulting combined potential has a unique center-symmetry preserving
minimum.%
\footnote
    {
    In the case of $\R \times T^3$, the undeformed theory was studied
    in detail by L\"uscher, van Baal and others in the mid 80's.
    (See the reviews \cite{Luscher:1998pe, vanBaal:2000zc} and
    references therein.)
    The goal of this ``QCD in a box" program was to use asymptotic freedom
    combined with the absence of the phase transitions in finite volumes
    to extract lessons about QCD on $\R^4$.
    This approach confines the theory to its short distance perturbative
    regime where it is perturbatively solvable.
    However, features of this small universe are very different from QCD
    on $\R^4$.
    In particular, when the physical box size becomes comparable to the inverse
    strong scale $\Lambda^{-1}$ there is a cross-over from a world of hadrons
    to a world of quarks and gluons.
    As noted in footnote \ref{thermo}, this cross-over becomes a
    sharp phase transition in the large $N$ limit.

    In our deformed Yang-Mills theory, as will be discussed below,
    there are no phase transitions as a function of box size
    even in the $N=\infty$ limit.
    For the deformed theory we will find that it is not
    $1/L$ which acts as an infrared cutoff, but rather $1/LN$.
    If $L N \Lambda \gg 1$ then the deformed theory reproduces
    the dynamics of Yang-Mills theory on $\R^4$.

    }

In the following sections we will, for simplicity, largely focus
on the case of just one compactified direction.

\subsection{Large $N$ equivalence between ordinary and deformed YM}

The most direct way to demonstrate equivalence between ordinary
Yang-Mills theory and our deformed theory is to compare the Schwinger-Dyson
(or loop) equations for gauge invariant observables.
As usual, for a rigorous treatment it is appropriate (and convenient)
to work with lattice regulated formulations of both theories.
It is also convenient to consider $U(N)$ gauge theories instead of $SU(N)$;
the difference in gauge groups only affects subleading $O(1/N^2)$
relative corrections to
Wilson loop expectation values or connected correlators.

Let $\delta^a_\ell$ denote an operator
which varies individual link fields according to
$\delta^a_\ell (U[\ell']) \equiv \delta_{\ell\ell'} \, t^a U[\ell]$,
where $\{ t^a \}$ is a set of $U(N)$ Lie algebra basis matrices
satisfying $\tr \> t^a t^b = \half \delta^{ab}$.
Invariance of the Haar measure implies that the integral of any
variation vanishes,
\begin{equation}
\int  d \mu_0 \; \delta^a_\ell \; ({\rm anything}) = 0 \,,
\end{equation}
where $d\mu_0 \equiv \prod_{\ell'} dU[\ell']$.
Choose (anything) to be $e^{S} \, \delta^a_\ell\,  W[C]$,
where $W[C] \equiv \frac 1N \tr\, U[C]$ is the Wilson loop
around some closed contour $C$.
Summing over the Lie algebra index $a$ and the link $\ell$ yields
\begin{equation}
\int  d\mu_0 \; e^{S}
\left( \delta S \cdot \delta W[C] + \delta^2 W[C] \right) =0 \,,
\end{equation}
where the dot product is shorthand for the sum over $a$ and $\ell$.
Dividing by the partition function $Z \equiv \int d\mu_0 \> e^S$
yields relations among expectation values,
\begin{equation}
 \big\langle \delta S  \cdot \delta W[C]  \big\rangle +
 \big\langle \delta^2 W[C] \big\rangle =0 \,.
\label{eq:loop1}
\end{equation}
These are Schwinger-Dyson equations for Wilson loop expectation values.

After working out the action of the variations,
the result may be expressed in a purely geometric form.
For lattice gauge theory with the Wilson action
one finds \cite{MM},
\begin{align}
    \half |C| \big\langle W[C]\big\rangle
    =
    &\sum_{\l \subset C} \;
    \sum_{p | \> \l\subset \partial p}
    \frac{\beta}{4N}
    \Big[
    \big\langle W[  (\overline {\partial p})  C]\big\rangle
    - \big\langle W[ ( {\partial p}) C]\big\rangle \Big]
\nonumber
\\
    &+ \sum_{\rm self-intersections} \mp \big\langle W[C'] W[C'']\big\rangle \,.
\end{align}
Here $|C|$ is the length of the loop $C$
({\em i.e.}, the number of links in the loop),
$W[(\partial p) C]$ denotes a Wilson loop which goes around
the boundary of plaquette $p$ (which contains a link contained in
the contour $C$) and then around the contour $C$,
and $\overline {\partial p}$ denotes the oppositely oriented
plaquette boundary.
The sum over self-intersections runs over all ways of decomposing
a loop $C$ which multiply traverses some link $\ell$
into two separate loops, $C = C'C''$, with the associated
sign determined by whether $C'$ and $C''$ traverse the link $\ell$
in the same or opposite directions.
See Ref.~\cite{KUY1} for more detailed discussion.

In the large $N$ limit,
with $\betatilde\equiv \frac{\beta} {N} = \frac{2}\lambda$ held fixed
(where $\lambda \equiv g^2 N$ is the 't Hooft coupling),
all $N$ dependence disappears.
Fluctuations in the values of Wilson loops vanish in this limit
(their distributions become arbitrarily sharply peaked).
This is a reflection of the classical nature of the large $N$ limit
\cite{LGY-largeN},
and implies that the expectation value of
a product of loops factorizes, up to $1/N^2$ corrections,
\begin{equation}
  \big\langle W[C'] W[C'']\big\rangle = \big\langle W[C']  \big\rangle \big\langle W[C'']\big\rangle
  + O(1/N^2) \,.
\end{equation}
(The $O(1/N^2)$ remainder is the connected correlator.)
Consequently,
in the large $N$ limit Wilson loop expectation values satisfy
a closed set of nonlinear algebraic equations,
\begin{eqnarray}
\half |C| \big\langle W[C]\big\rangle =  \sum_{\l \subset C} \sum_{p| l\subset \partial p}
\frac{\betatilde}{4} \,
\Big[
\big\langle W[  (\overline {\partial p})  C]\big\rangle - \big\langle W[  ( {\partial p})  C]\big\rangle \Big]  \cr
+ \sum_{\rm self-intersections}
\mp  \big\langle W[C'] \big\rangle \> \big\langle W[C'']\big\rangle \,.
\end{eqnarray}

The loop equations in the concise form (\ref {eq:loop1})
are equally valid for the deformed theory.
The only difference is that $S$ now includes the double trace
deformation $\Delta S$, and this generates new terms
in the loop equations
given by $\langle \delta (\Delta S) \cdot \delta W[C] \rangle$.
Just as the usual Wilson action leads to terms
in which a plaquette is inserted into the loop $C$,
the piece of $\Delta S$ proportional to $|\tr\, \Omega^k|^2$
generates terms in the loop equation
in which the topologically non-trivial loop $\Omega^k$ (or its inverse)
is ``sewn'' into the loop $C$
(if $C$ contains links pointing in the compactified direction).
But because $\Delta S$ contains absolute squares of traces,
each such term is multiplied by the complex conjugate
of the trace of the inserted loop.
Hence,
\begin{equation}
  \big\langle \delta (\Delta S) \cdot \delta W[C] \big\rangle
  =
  \sum_{k \ne 0}
  b_k[C] \>
  \big\langle W[\Omega^k C] \, W[\Omega^{-k}] \big\rangle \,,
\end{equation}
where $\Omega^k C$ denotes a loop obtained by concatenating $\Omega^k$
and $C$ at their intersection links,
and the coefficients $b_k[C]$ are proportional to $a_{|k|}$ but also
depend on the number of links in $C$ which point in the compactified direction.
The essential point is that the variation acts on one of the
two traces comprising the double trace deformation, leaving the other
trace unchanged.
In the large $N$ limit, due to factorization,
\begin{equation}
     \big\langle W[\Omega^k C] W[\Omega^{-k}] \big\rangle
     = \big\langle W[\Omega^k C] \big\rangle \, \big\langle  W[\Omega^{-k}] \big\rangle
     +  O(1/N^2) \,.
\end{equation}
But for any $k$ which is non-zero modulo $N$,
$W[\Omega^{-k}]$
transforms non-trivially (acquiring a phase $e^{-2 \pi i k/N}$)
under a $\Z_N$ center symmetry transformation.
Hence its expectation value
is an order parameter for the center symmetry
and $\langle W[\Omega^{-k}] \rangle$ must vanish
in any phase with unbroken center symmetry.
As discussed above,
the deformed theory is constructed so as to ensure unbroken
center symmetry for all compactification radii.
Consequently,
all additional terms in the loop equations generated by the
deformation of the action vanish in the large $N$ limit,
\begin{equation}
    \big\langle \delta (\Delta S) \cdot \delta W[C] \big\rangle = O(1/N^2) \,,
\end{equation}
implying that Wilson loops in the original and deformed Yang-Mills theory
satisfy identical large-$N$ Schwinger-Dyson equations.

Ordinary Yang-Mills theory has unbroken center symmetry only
for sufficiently large compactifications, $L > L_{\rm c}$.
The coinciding large $N$ loop equations in ordinary and deformed
Yang-Mills theories imply that Wilson loop expectation values in
these two theories have identical large $N$ limits
when $L > L_{\rm c}$.%
\footnote
    {
    This argument, that coinciding loop equations imply coinciding
    expectation values, glosses over the possibility that the
    infinite set of loop equations may have multiple solutions
    which respect center symmetry, with different theories
    potentially corresponding to different solutions of the
    same set of equations.
    The alternative approach of comparing the $N = \infty$ classical dynamics
    generated by appropriate large $N$ coherent states,
    discussed in Ref.~\cite{KUY2},
    eliminates this loophole
    and demonstrates equivalence in any phase of the theories
    which satisfy the necessary and sufficient
    symmetry realization conditions.
    }

The same approach may be used to compare the Schwinger-Dyson
equations satisfied by connected correlators of two or more Wilson loops,
with exactly the same conclusion: the leading large $N$ behavior
of connected correlators coincide between ordinary and deformed
Yang-Mills theories, provided $L > L_{\rm c}$.
Thus, in sufficiently large volume
the net effect of the double trace deformation on the dynamics
of the theory is $\O(1/N^2)$,
and vanishes in the large $N$ limit.
In another words, the physics of the deformed Yang-Mills theory depends
on the deformation parameters $\{ a_i \}$ only in the combination
$a_i/N^2$ which vanishes at $N=\infty$.
This demonstrates the nonperturbative equivalence of ordinary Yang-Mills theory
and the deformed YM theory, formulated on $\R^3 \times S^1$
(or more generally, on any toroidal compactification of flat space),
provided the compactification size is above the critical size for
center symmetry breaking in the undeformed theory.

\subsection{Large $N$ volume independence of deformed YM theory}

Unbroken center symmetry is
necessary and sufficient for the validity of the
large $N$ volume independence of Yang-Mills theory
(or more general gauge theories containing adjoint
representation matter fields).
This may be demonstrated by comparing large $N$ loop equations,
or the $N = \infty$ classical dynamics generated by suitable
coherent states \cite{Kovtun:2007py}.
Corrections to this equivalence for finite $N$ scale as $1/N^2$.
The loop equation analysis is very similar to that sketched above.
In the large $N$ loop equations for topologically trivial Wilson loops,
one finds that the only volume-dependent terms (arising from
self-intersections) automatically vanish as long as the
center symmetry is not spontaneously broken.

The analysis of large $N$ volume independence in Ref.~\cite{Kovtun:2007py}
applies equally well to the deformed theory which, by construction,
has unbroken center symmetry for any compactification radius.
Because the double trace operators in $P[\Omega]$ are squares
of loops with non-zero $N$-ality,
the presence of the deformation potential $P[\Omega]$
has no effect on the large $N$ classical dynamics
of center-symmetry symmetric states.
Consequently, deformed Yang-Mills theory, in the large $N$ limit,
is completely volume independent.

In the lattice formulation, if one compactifies all directions
then one may reduce the lattice size all the way down to a single site,
in which case the theory becomes a simple matrix model
of Wilson lines $\{ \Omega_i \}$ running in each lattice direction
with action,
\begin{equation}
    S^{\rm deformed}_{\rm single-site}
    =
    \frac\beta 2
    \sum_{i>j=1}^d
    \tr \,
    \bigl(
    \Omega_i \Omega_j \Omega_i^\dagger \Omega_j^\dagger
    +
    \Omega_j \Omega_i \Omega_j^\dagger \Omega_i^\dagger
    \bigr)
    +
    P[\Omega_1,\cdots,\Omega_d] \,.
\label{eq:one-site}
\end{equation}
The large $N$ limit of this matrix model will reproduce the leading large
$N$ behavior of expectation values and connected correlators of
Wilson loops in uncompactified Yang-Mills theory.
As discussed in the Introduction, the single-site deformed Yang-Mills
theory (\ref{eq:one-site}) provides a simple generalization of Eguchi-Kawai
reduction which is valid for any value of the lattice coupling $\beta$.

\subsection {Addition of matter fields}

Consider adding $\Nf$ species of matter fields (either fermions or scalars)
in the fundamental representation to $SU(N)$ Yang-Mills theory,
either ordinary or deformed, with one dimension compactified.
The addition of fundamental representation matter explicitly breaks
the $\Z_N$ center symmetry.
However, if $\Nf$ is held fixed as $N \to \infty$,
then the fundamental representation matter fields have only
a subleading $O(\Nf/N)$ effect on the gauge field dynamics.
As a result, everything discussed above remains valid.
That is, the leading large $N$ behavior of
expectation values or connected correlators of Wilson loops in
the undeformed theory, in sufficiently large volume, coincide
with the corresponding observables in
the theory, in arbitrary volume,
deformed by the addition of the stabilizing potential $P[\Omega]$.
In addition, one may also show that the same equivalence applies
to the leading large $N$ behavior of mesonic expectation values
and connected correlators.%
\footnote
    {
    This is easiest to understand by considering the equivalent
    gluonic observables produced by integrating out the matter fields.
    For the case of large $N$ volume independence in the undeformed
    theory, see Ref.~\cite{Kovtun:2007py} for details.
    The presence of the deformation potential does not affect this analysis.
    }
Note, however, that these large $N$ equivalences cease to apply
if $\Nf/N$ is held fixed as $N \to \infty$ \cite{Kovtun:2007py}.

If adjoint representation matter fields are added to the theory
(ordinary or deformed), then the large $N$ equivalences discussed above
also remain valid.
Adding adjoint representation fields enlarges the natural set of gauge
invariant observables from simple Wilson loops to Wilson loops decorated
by arbitrary numbers of insertions of adjoint matter fields.
But the presence of adjoint matter fields
preserves the center symmetry of the underlying Yang-Mills theory.
As a result, the above-described comparison of large $N$ loop equations
(or large $N$ classical dynamics)
between the ordinary and deformed theories
immediately generalizes to the case of Yang-Mills
theories with adjoint matter, with exactly the same conclusions.%
\footnote
    {
    For a detailed discussion of loop equations in theories
    with adjoint matter, see Ref.~\cite{KUY1}.
    }

Finally, one may also consider the addition of
matter fields in rank-two antisymmetric or symmetric tensor
representations
(yielding theories we will refer to as QCD(AS) or QCD(S),
respectively).
The presence of fields in these representations
reduces the $U(1)$ center symmetry of $U(N)$ Yang-Mills theory
down to $\Z_2$.
Given the central role the center symmetry played in the
above large $N$ equivalences,
one might think this reduction in center symmetry would destroy
these large $N$ equivalences.
This is not the case.
One way to see this is to note,
as discussed in Ref.~\cite{Kovtun:2007py},
that volume-dependent terms in the $N = \infty$
loop equations only appear if
loops with non-zero winding number around the compactified direction
acquire non-zero expectation values.
The addition of the deformation potential $P[\Omega]$
prevents topologically non-trivial Wilson loops from acquiring
non-zero large $N$ expectation values, even in small volumes.
Consequently, the situation is analogous to the pure gauge case:
the leading large $N$ behavior of
expectation values or connected correlators of single trace
observables in QCD(AS/S) in sufficiently large volume
coincide with the corresponding observables in
the theory modified by the addition of the deformation potential
in arbitrary volume, as depicted in Fig.\ref{fig:mappings}.%
\footnote
    {
    Another way to understand this is to note the existence of a large $N$
    equivalence (so-called ``orientifold equivalence'') between
    theories with rank-two symmetric or antisymmetric representation
    matter and corresponding theories with adjoint representation matter
    [``QCD(adj)''] \cite{UY,ASV1, Armoni:2004ub}.
    This large $N$ equivalence applies to the charge-conjugation even
    sectors of the two theories, and only holds if charge conjugation
    symmetry is not spontaneously broken.
    When, for example, the matter fields are fermions
    with periodic boundary conditions,
    examination of the Wilson line effective potential shows that
    QCD(AS/S) {\em does\/} spontaneously break both charge conjugation and
    center symmetry when compactified with sufficiently small size \cite{UY}.
    But the addition of a deformation potential of the form
    (\ref{eq:P[Omega]}) (with sufficiently positive coefficients)
    will prevent this spontaneous symmetry breaking, just as it does
    in the pure Yang-Mills case.
    Since QCD(adj) satisfies large $N$ volume independence
    (as long as its center symmetry is not spontaneously broken),
    the same large $N$ volume independence must also apply to QCD(AS/S)
    (as long as charge conjugation is not broken).
    In sufficiently large volumes, there is no reason to believe
    that charge conjugation symmetry breaks spontaneously in
    QCD-like theories with rank-two tensor representation matter.
    Therefore, large $N$ orientifold equivalence combines with
    large $N$ volume independence of QCD(adj) to imply
    volume independence in QCD(AS/S) as long as center and charge
    conjugation symmetries are not spontaneously broken --- which
    is what the deformation potential ensures.
    }

\section {Confinement at small radius and small $N$}

When compactified on a small circle, $L \ll \Lambda^{-1}$,
the gauge coupling of the deformed theory
is small at the compactification scale, $g^2 ({1}/{L}) \ll 1$.
As discussed earlier,
the combined potential $\mathcal V[\Omega] + P[\Omega]$ is minimized when
\begin{equation}
    \Omega =
    {\rm Diag} \left(
	1, e^{2\pi i/N}, e^{4\pi i/N}, \cdots , e^{2\pi i(N-1)/N}
    \right) ,
\label{eq:vev}
\end{equation}
up to conjugation by an arbitrary $SU(N)$ matrix.
Working in a gauge in which $\Omega$ is diagonal,
and using (temporarily) gauge-dependent language,
this configuration may be regarded as breaking the gauge symmetry
down to the maximal Abelian subgroup,
\begin{equation}
    SU(N) \rightarrow  U(1)^{(N-1)} \,.
\end{equation}
Modes of the diagonal components of the $SU(N)$ gauge field
with no momentum along the compactified $\hat x_4$ direction
describe photons associated with the Cartan subgroup of $SU(N)$.
Modes of the diagonal components of the gauge field with non-zero
momentum in the compactified direction form a Kaluza-Klein tower and
receive masses which are integer multiples of $2\pi/L$.
The off-diagonal components of the $SU(N)$ gauge field
describe Kaluza-Klein towers of $W$-bosons which are charged
under the unbroken $U(1)^{N-1}$ gauge group.
The non-zero value of $A_4 \equiv -(i/L) \ln \Omega$ shifts the masses
of these off-diagonal components by multiples of $2\pi/(NL)$.
The net effect is that there are charged $W$-bosons with masses
\begin{equation}
    m_{W_k}= \frac{2\pi \, k}{NL} \,,
    \qquad  k=1, 2, \cdots, \,.
\label{eq:m_W_k}
\end{equation}
For later convenience, we define $m_W$ to be
the mass of the lightest $W$ bosons,
\begin{equation}
    m_W \equiv \frac{2 \pi}{NL} \,.
\end{equation}
This is the mass scale below which the dynamics is effectively Abelian.%
\footnote
    {
    Fluctuations in the eigenvalues of $\Omega$
    away from the minimum (\ref{eq:vev})
    correspond to neutral ``Higgs bosons''.
    The masses of these fluctuations
    depend on the coefficients
    $\{ a_n \}$ of the deformation potential but
    parametrically are of order
    $\sqrt \lambda/L \sim \sqrt \lambda \, N  \, m_W $.
    This scale will be large compared to the mass scale of the
    non-perturbative $3d$ dynamics,
    and these fluctuations will play no role in the following discussion.
    }
Note that, at fixed $L$, the lightest $W$ bosons have masses
which become small when $N \to \infty$.
This will be important in the discussion of the large $N$ behavior
of the deformed YM theory.
But first, in this section, we consider the dynamics of the deformed
theory when $N$ is fixed and small.

The $N{-}1$ photons of the Cartan subgroup do not couple (directly)
to the Wilson line and remain massless to all orders in
perturbation theory.
Thus, a strictly perturbative analysis would lead one to expect
that the deformed theory, for sufficiently small $L$,
would have a non-confining Coulomb phase.
We will see that this is incorrect --- nonperturbative effects
lead to the generation of a mass gap and produce confining
long distance physics.

The analysis of non-perturbative properties in our compactified
deformed Yang-Mills theory is very similar to Polyakov's treatment
of the $3d$ Georgi-Glashow model \cite{Polyakov:1976fu}.
But instead of a three-dimensional theory with a non-compact
Higgs field, we have a compactified four-dimensional theory with
the group-valued Wilson line $\Omega$ serving as a compact Higgs field.
For theories involving massless complex fermions,
the difference between compact and noncompact
Higgs systems can be major \cite{Unsal:2008}.
However, in our case, the differences relative to
Polyakov's classic discussion are rather minimal.

Due to the $SU(N) \rightarrow U(1)^{N-1}$ gauge symmetry ``breaking'',
there exist topologically stable, semiclassical
field configurations, namely monopoles \cite{Gross:1980br}.
At the center of a monopole, one pair of eigenvalues of the
Wilson line become degenerate.
For fundamental ({\em i.e.}, minimal action) monopoles, 
this will be a pair of eigenvalues which are nearest-neighbors at infinity.
If the adjoint Higgs field was noncompact, then there would
be $N{-}1$ species of fundamental monopoles.
This follows from the topological considerations:
the second homotopy group
$
    \pi_2[SU(N)/U(1)^{N-1}]
    =
    \pi_1[U(1)^{N-1}]
    =
    \Z^{N-1}
$,
implying that fundamental monopoles come in $N{-}1$ varieties.
However,
with a compact Higgs field
there is an extra fundamental (``Kaluza-Klein'') monopole which arises due to
the fact that the underlying  theory is formulated  on a cylinder,
$\R^3 \times S^1$, or equivalently that the configuration space
of $\Omega$ is compact.

The monopoles may be characterized by their magnetic charges,
topological charge, and their action.
The magnetic charges of the  $N$ different types of fundamental monopoles
are proportional to the simple roots and affine root of the Lie algebra
of the unbroken $U(1)^N$ gauge group.%
\footnote
    {
    Even though the gauge symmetry ``breaking'' is
    $SU(N) \rightarrow U(1)^{N-1}$,
    for ease of presentation it is convenient to add an extra photon to
    the original theory and discuss $U(N) \rightarrow U(1)^N$.
    This simplifies the discussion of charge assignments of monopoles,
    and the affine roots of the associated Lie algebra.
    In the continuum limit of the theory this extra photon
    completely decouples from the other degrees of freedom
    and may simply be ignored.
    It should not be confused with the $N{-}1$ photons which
    have non-trivial nonperturbative dynamics.
    }
The simple roots are given by%
\footnote
    {
    This set of simple roots corresponds to choosing
    Lie algebra generators normalized to satisfy
    $\tr \> t^a t^b = \delta^{ab}$,
    instead of $\half \, \delta^{ab}$ as in the previous section.
    }
\begin{subequations}
\begin{align}
    \alpha_1&= (1, -1, 0, \ldots, 0)= \hat e_1-\hat e_2 \,,\\
    \alpha_2&= (0, 1, -1, , \ldots, 0)=\hat e_2 -\hat e_3  \,,\\
    &\;\; \vdots   \nonumber\\
    \alpha_{N-1}&= (0, \ldots, 0,  1, -1)=\hat e_{N-1} -\hat e_{N} \,,
\end{align}
\label{eq:roots}
\end{subequations}
 and the affine root is
\begin{equation}
    \alpha_N \equiv  - \sum_{j=1}^{N-1} \,\alpha_j
    =  (-1, 0, 0, \ldots ,1)= \hat e_N- \hat e_1 \,.
\end{equation}
For later convenience,  let $\Delta^{0}_{\rm aff}$
denote the affine (extended) root system of the the associated Lie algebra,
\begin{equation}
    \Delta^{0}_{\rm aff} \equiv
    \{ \alpha_1, \alpha_2, \ldots , \alpha_{N-1},  \alpha_N \} \,.
\end{equation}
It is the affine root system which is
relevant for compact Yang-Mills Higgs systems.
The roots $\alpha_i \in \Delta^{0}_{\rm aff}$ obey
\begin{equation}
    \alpha_i \cdot \alpha_j
    = 2 \delta_{i,j}
    - \delta_{i, j + 1}
    - \delta_{i, j - 1}
    \,, \qquad i, j=1, \ldots N \,.
\label{Eq:inner}
\end{equation}
The form (\ref{Eq:inner}) of these inner products
will translate into self and nearest neighbor interactions
between monopoles in the Dynkin space.
The above choice of basis is natural due its visual simplicity,
but the inner products (\ref{Eq:inner}) of the roots of the
associated Lie algebra are basis independent.

Let $F^\mu \equiv \frac 1{2g} \, \epsilon^{\mu\nu\lambda 4} \, F_{\nu\lambda}$
denote the $U(1)^{N}$ valued $3d$ magnetic field,
with conventional perturbative normalization.
(In a gauge where $\Omega$ is diagonal, $F^\mu$ is just the list of
diagonal elements of the original non-Abelian field strength,
multiplied by $1/g$.)
The magnetic charges of a monopole of type $i = 1, \cdots, N$
are given by the
root $\alpha_i$ (up to a factor of $2\pi/g$),
\begin{equation}
 \int_{S^2} d{\bm\Sigma} \cdot \mathbf F = \frac {2 \pi}{g} \,\alpha_i
 \qquad
 \qquad
 \mbox{[type $(i)$ monopole]}
 \,.
\label{eq:mag-charge}
\end{equation}
(The $S^2$ is an arbitrarily large sphere in $\R^3$.
The flux is independent of the value of $x^4$ at which
the integral is performed, as the long distance monopole fields
are independent of $x^4$.)

The topological charge is correlated with the magnetic charge
of the monopole.
For fundamental monopoles
with magnetic charges $\alpha_i \in \Delta^0_{\rm aff}$,
the topological charge is
\begin{equation}
    \nu \equiv
    \int_{\R^3 \times S^1} \frac{1}{16 \pi^2} \>
    \tr\, F_{\mu \nu} {\widetilde F}^{\mu \nu}
    =
    \frac{1}{N} \,.
\end{equation}
For antimonopoles with magnetic charges
$-\alpha_i$, the topological charge $\nu = -1/N$.

The electric charges of $W$-bosons may also be
simply expressed in terms of the affine roots $\Delta_{\rm aff}^{0}$.
The lightest $W$ bosons, with mass $m_W$,
may be labeled by a single root which gives their electric
charges (up to a factor of $g$),
\begin{equation}
    Q_{W_{\alpha_i}} = g \, \alpha_i \,.
\end{equation}
$W$-bosons in the next heavier multiplet
are labeled by a pair of neighboring roots, and have charges
\begin{equation}
    Q_{W_{\alpha_i+ \alpha_{i+1}}} = g\, (\alpha_i + \alpha_{i+1}) \,,
\end{equation}
{\em etc}.
Dot products of the $W$-boson charges
and monopole charges obey the Dirac quantization condition,
\begin{equation}
    Q_{W_{\alpha_i}} \cdot Q_{M_{\alpha_j}}
    = g \, \alpha_i \cdot \frac{2 \pi} {g} \, \alpha_j
    = 2\pi ( 2\delta_{ij} - \delta_{i,j+ 1} - \delta_{i,j- 1})
    = \left\{ \begin{array}{rl}
	    4 \pi \,, & \qquad {\rm for} \; i=j \,;  \cr
	    -2 \pi\,, & \qquad {\rm for} \; i=j \pm1 \,;       \cr
	    0\,,~ & \qquad {\rm otherwise}.
	\end{array}
       \right.
\end{equation}

Conjugation by a $\Z_N$ ``shift'' matrix,
which is part of the global gauge symmetry,
cyclically permutes the Wilson line eigenvalues and hence
cyclically permutes the $N$ different species of
fundamental monopoles.
The presence of this symmetry (which is one of the features which
distinguishes compact and non-compact Higgs systems)
guarantees that the $N$ different
types of fundamental monopoles have identical values of the action.
Monopole solutions are self-dual,
\begin{equation}
 F_{\mu\nu}=  \widetilde F_{\mu\nu} \, , 
\end{equation}
and hence the Yang-Mills action
of a fundamental monopole (or antimonopole) is
\begin{equation}
    S_{\rm YM}
    =
    \int_{\R^3\times S^1} \frac{1}{2g^2} \> \tr \, F_{\mu\nu}^2
    =
    \left|
	\int_{\R^3\times S^1} \frac{1}{2g^2} \>
	\tr \, F_{\mu\nu} \widetilde F^{\mu\nu}
    \right|
    =
    \frac{8\pi^2}{g^2} \> |\nu|
    =
    \frac{8 \pi^2}{g^2 N} \,.
\end{equation}
After adding the contributions of the deformation potential $P[\Omega]$
and the induced one-loop effective potential $\mathcal V[\Omega]$,
the complete monopole action will differ from this value.
But the deviation is perturbative in $g^2$,
so the monopole action
\begin{equation}
    S_0 \equiv S_{\rm YM} + \Delta S + S_{\rm 1-loop}
    =\frac {8\pi^2}{g^2 N} + O(1) \,.
\label{eq:S0}
\end{equation}

The correct infrared description of the deformed Yang-Mills theory
on $\R^3 \times S^1$ at small radius
is generated by a dilute gas of monopoles (and antimonopoles)
of $N$ different types,
interacting via the species-dependent long range Coulomb potential,
\begin{equation}
    V_{\pm i,\pm j}(\mathbf r)
    =
    L\left(\frac{2 \pi}{g}\right)^2
    \frac{(\pm \alpha_i) \cdot (\pm \alpha_{j}) }{4 \pi |\mathbf r|}
    =
    \pm L\left(\frac{2 \pi}{g}\right)^2 \,
    \frac{2 \delta_{ij} - \delta_{i, j+1} - \delta_{i, j-1}}{4 \pi |\mathbf r|}
    \,.
\label{Eq:Dynkin}
\end{equation}
(The overall sign is plus for monopole-monopole,
and minus for monopole-antimonopole.)
Hence, we are dealing with a multi-component classical
plasma, with nearest-neighbor interactions in the Dynkin space.
As with any classical plasma, this system will exhibit Debye screening.
The field due to a static external magnetic charge will fall exponentially
with distance,
$|\mathbf F| \sim e^{- m_D r}/r$,
with $m_D^{-1}$ the characteristic Debye screening length.
This implies that external fields cannot propagate coherently over
distances large compared to the Debye length, which will be the
longest correlation length in the system.
The Debye mass $m_D$ will appear as a dynamically generated photon mass.
This will be shown explicitly.

For momentum scales small compared to the lightest $W$ mass, 
the equilibrium dynamics is correctly represented by a
grand canonical ensemble of all types of monopoles and antimonopoles.
Consider a configuration in which there are $n^{(i)}$ monopoles
and $\bar n^{(i)}$ antimonopoles of types $i = 1,\cdots,N$,
located at positions $\r^{(i)}_k$, $k = 1, \cdots, n^{(i)}$
and $\bar {\r}^{(i)}_l$, $l = 1, \cdots, \bar n^{(i)}$,
respectively.
The magnetic field generated by this ensemble of magnetic charges is
\begin{eqnarray}
    {\mathbf B}(\x)
    =  \sum_{i=1}^{N}
    \frac{2 \pi}{g} \, \alpha_i
     \left[
     \sum_{k=1}^{n^{(i)}}
    \frac{\x- \r^{(i)}_k}{4 \pi |\x- \r^{(i)}_k |^3}
    -
     \sum_{l=1}^{\bar n^{(i)}}
    \frac{\x- \bar \r^{(i)}_l}{4 \pi  |\x- \bar \r^{(i)}_l |^3}
    \right] .
\end{eqnarray}
The action of such a monopole configuration is the sum of the monopole
self-energies plus their potential energy due to Coulomb interactions,
\begin{equation}
    S_{\rm monopole-gas}
    =
    S_0 \sum_{i=1}^{N} \> \left( n^{(i)} + \bar n^{(i)}\right)
    +  S_{\rm int} \,,
\end{equation}
with
\begin{equation}
    S_{\rm int}
    =
    \frac {2\pi^2L}{g^2}
    \sum_{i, j=1 }^{N}
    \alpha_i \cdot \alpha_j
    \left[
	\sum_{k=1}^{n^{(i)}}
	\sum_{l=1}^{n^{(j)}} G(\r^{(i)}_k{-}\r^{(j)}_{l})
	+
	\sum_{k=1}^{\bar n^{(i)}}
	\sum_{l=1}^{\bar n^{(j)}} G(\bar\r^{(i)}_k{-}\bar\r^{(j)}_{l})
	-2
	\sum_{k=1}^{n^{(i)}}
	\sum_{l=1}^{\bar n^{(j)}} G(\r^{(i)}_k{-}\bar\r^{(j)}_{l})
    \right] ,
\label{Eq:Dynkin2}
\end{equation}
and
\begin{equation}
    G(\r)  \equiv
    \frac{1}{4 \pi |\r|} \,.
\label{eq:G}
\end{equation}

The grand canonical partition function of this multi-component Coulomb gas is
\begin{equation}
    Z =
    \prod_{i=1}^N
    \left\{
    \sum_{n^{(i)}=0}^\infty
    \frac {\zeta^{n^{(i)}}}{n^{(i)}!}
    \sum_{\bar n^{(i)}=0}^\infty
    \frac {\zeta^{\bar n^{(i)}}}{\bar n^{(i)}!}
    \int_{\R^3}
    \prod_{k=1}^{n^{(i)}} d\r^{(i)}_k
    \int_{\R^3}
    \prod_{l=1}^{\bar n^{(i)}} d\bar\r^{(i)}_l
    \right\} \;
    e^{-S_{\rm int}} \,,
\label{eq:grand}
\end{equation}
where
\begin{equation}
    \zeta \equiv
    C \, e^{-S_0}
    = A \,  m_W^3 \, (g^2N)^{-2} \, e^{-\Delta S} \, e^{-8\pi^2/N g^2(m_W)}
 \label{eq:fug}
\end{equation}
is the monopole fugacity.
The prefactor $C$ represents the one-loop functional determinant in
the monopole background.
Extracting the zero-modes of the small fluctuation operator via the
usual collective coordinate procedure
leads to factors of $(g^2N)^{-2} \, m_W^3$.
(See the appendix for details.)
If the coupling is evaluated at the scale $m_W$,
which is natural for this problem,
then the non-zero mode part of the one-loop determinant
merely gives rise to an overall
dimensionless (and $N$ independent) coefficient $A$.
In the final form of (\ref{eq:fug}),
$\Delta S$ denotes the deformation term in the action (\ref{eq:Delta S})
evaluated in the background of a fundamental monopole.
This is an $O(1)$ number, independent of the coupling $g^2$,
whose explicit value depends, of course,
on the deformation parameters $\{ a_n \}$.
[For large $N$, $\Delta S$ scales as $O(1/N)$.]

Using the fact that $G(\r)$ is the Green's function for the $3d$ Laplacian,
this partition function can be exactly transformed into a $3d$ scalar
field theory with an $N$-component real scalar field,
\begin{equation}
    Z =
    \int \prod_{i=1}^N \mathcal D\sigma_i \;
    e^{-S_{\rm dual}[\bm\sigma]} \,,
\label{eq:Zdual}
\end{equation}
where
\begin{equation}
    S^{\rm dual}
    =
    \int_{\R^3} \;
    \Big[
	\frac{1}{2L}
	\left(\frac{g}{2\pi}\right)^2
	(\nabla \bm\sigma)^2
	-
	\zeta \>
	\sum_{i=1}^N \>
	\cos (\alpha_i \cdot \bm\sigma)
    \Big] \,.
\label{eq:Sdual}
\end{equation}
To verify this, it is easiest to start with the functional integral
(\ref {eq:Zdual}),
rewrite the cosines in terms of exponentials of $\bm\sigma$,
expand the exponential of each of the resulting interaction terms
in a power-series in $e^{-S_0} \, e^{\pm i\alpha_i \cdot \bm\sigma}$,
and then perform the functional integral over $\bm\sigma$.
The scalar fields $\sigma_i$ appearing in this representation
are dual fields for the $3d$ Abelian gauge fields $A^i_\mu$.%
\footnote
    {
    In three dimensions, Abelian duality relates a photon to a compact scalar.
    With $\sigma^j(\x) $ the compact scalar dual to the
    photon $A^{(j)}_{\mu}(\x)$ of the $j$'th $U(1)$ subgroup,
    the Abelian duality relations are
    \begin{equation*}
	* d \sigma^j =  \half L  \; \Im (\tau) \,  F^{(j)} \,,\qquad
	\tau(L^{-1})= \frac{ 4\pi i}{g^2} + \frac{\theta}{2 \pi} \,,\qquad
	F_{\mu \nu}^{(j)} = \frac{g^2} {2 \pi  L} \>
	\epsilon_{\mu \nu \rho} \, \partial_{\rho} \sigma^j \,.
    \end{equation*}
    The $3d$ Maxwell action becomes
    $
	\frac{L}{4g^2} (F^{(j)}_{\mu \nu})^2
	=
	\frac {L \Im(\tau)}{16\pi} \, (F^{(j)}_{\mu \nu} )^2
	=
	\frac 1{2\pi L \Im( \tau)} \,
	(\partial_\mu \sigma^j)^2
    $.
    The path integral of the Abelian gauge theory in the presence of a monopole
    with charge $\pm \alpha_j$ located at position $\x$ is equivalent
    to the insertion of $e^{ \pm i \alpha_j\cdot\sigma(\x)}$
    into the path integral over the dual scalar fields
    \cite{Polyakov:1976fu,Deligne:1999qp}.
    The complete partition function of the long distance effective theory is
    a sum over all topological sectors,
    each of which may contain an arbitrary number of monopoles
    and antimonopoles
    (whose charges sum to give the appropriate topological class).
    Summing over all numbers and locations of monopoles (and antimonopoles),
    weighted with the appropriate fugacity,
    directly yields the result (\ref{eq:Sdual}).
    }

The fields $\{ \sigma_i \}$ should be regarded as compact scalar fields
defined modulo $2\pi$.
In addition to invariance under $2\pi$ shifts in any component of $\bm\sigma$,
note that the monopole induced interaction vertex has the additional
shift symmetry 
\begin{equation} 
    \bm \sigma \rightarrow \bm \sigma + 2 \pi \mu_i, \qquad i= 1, \ldots, N{-}1
\label{eq:shiftsym}
\end{equation}
where $\{ \mu_i \}$ are the $N{-}1$
fundamental weights of the $SU(N)$ algebra.
These are defined by the reciprocity relation with the simple roots,
\begin{equation}
    \mu_i \cdot \alpha_j
    = \half \delta_{ij} \, \alpha_j^2 
    = \delta_{ij}
\label{eq:reciprocity}
\end{equation}
for $i=1, \cdots, N{-}1$,
which implies that the fundamental weights $\{ \mu_i \}$
form a basis which is dual to the fundamental roots $\{ \alpha_j \}$.
The presence of the symmetry (\ref{eq:shiftsym})
is related to the fact that the vacuum of the original theory
can be probed by $N{-}1$ different types of external charges,
distinguished by their (non-zero) values of $N$-ality.
This will be discussed below.

Including a non-zero theta parameter in the original Yang-Mills
action,
\begin{equation}
    S^{\rm YM}
    \to
    S^{\rm YM}
    +
    i \theta
    \int_{\R^3 \times S^1} \frac{1}{16\pi^2} \>
    \tr\, F_{\mu \nu} {\widetilde F}^{\mu \nu} \,,
\end{equation}
has the effect,
in the grand canonical partition function (\ref {eq:grand}),
of multiplying the monopole fugacity by $e^{i\theta/N}$
and antimonopole fugacity by $e^{-i\theta/N}$.
In the dual representation (\ref {eq:Sdual}), this amounts to
shifting the argument of the cosine by $\theta/N$,
so that the interaction becomes
\begin{equation}
    -\zeta \sum_{i=1}^N \cos(\alpha_i \cdot \bm \sigma + i\theta/N) \,.
\end{equation}
In this form, $2\pi$ periodicity of the theory with respect to $\theta$
is not manifest.
However, a shift of the dual scalar fields,
$
    \bm \sigma \to \bm \sigma + (\theta/N) \beta
$
with $\beta \equiv (0,1,2, \cdots, N{-}1)$,
converts the interaction term to the manifestly $2\pi$ periodic form
\begin{equation}
    -\zeta \left[
    \sum_{i=1}^{N-1} \cos(\alpha_i \cdot \bm \sigma)
    +
    \cos(\alpha_{N} \cdot \bm \sigma + \theta)
    \right] ,
\end{equation}
in which theta dependence only appears in the term involving the
affine root.
For simplicity,
in the following subsections we will focus on the case of $\theta = 0$.

\subsection {Mass gap}

The cosine potential in the dual action (\ref {eq:Sdual})
generates a mass term for the photons.
Rescaling $\bm\sigma$ to put the kinetic term into canonical form
and expanding the potential to  quadratic order around the minimum
at $\bm \sigma = 0$
gives
\begin{equation}
    V(\sigma_i) =
    (\mbox{const.}) +
    \half  m_\gamma^2 \, \sum_{i=1}^{N} \> (\sigma_{i+1}- \sigma_i)^2 \,,
\end{equation}
with $\sigma_{N+1} \equiv \sigma_1$ and
\begin{equation}
    m_\gamma^2
    \equiv  \frac {(2\pi)^2}{g^2} \, L\, \zeta
    =  A \,  m_W^2 \left(\frac {2 \pi}{g^2 N}\right)^3 \,
    e^{-\Delta S} \,
    e^{-8\pi^2/(N g^2(m_W))} \,.
\end{equation}
A $\Z_N$ Fourier transform,
\begin{equation}
    \tilde\sigma_p \equiv
    \frac{1}{\sqrt N} \sum_{j=1}^{N} \> e^{-2\pi i p j/N} \, \sigma_j \,,
    \qquad p = 0, \cdots, N{-}1 \,,
\label{Eq:Fourier}
\end{equation}
diagonalizes this mass term and yields
\begin{equation}
    V(\sigma_i) =
    (\mbox{const.}) +
    \half  \sum_{p=0}^{N-1} \> m_p^2 \> |\tilde\sigma_p|^2 \,,
\end{equation}
with
\begin{equation}
    m_p \equiv m_\gamma \, \sin \frac{\pi p }{N} \,.
\label{eq:mp}
\end{equation}
Expressing $m_\gamma$ in terms of the renormalization group invariant scale
$\Lambda$, defined by
\begin{equation}
    \Lambda^{b_0} =
    \mu^{b_0} \, \left({Ng^2(\mu)}\right)^{-b_1/b_0} \,
    e^{-8\pi^2/( \, Ng^2(\mu)) }
    \label{eq:RG}
\end{equation}
(with $b_0 = 11/3$ and $b_1 = 17/3$),
yields
\begin{equation}
    m_\gamma
    ={\widetilde A}  \; \Lambda \; 
        (\Lambda NL)^{5/6} \,
    \left| \ln N\Lambda L \right|^{9/11}
    \,,
\label{eq:mgamma}
\end{equation}
where ${\widetilde A}$ is an $O(1)$ coefficient.
Relative corrections suppressed by powers of
$g^2(m_W) \sim 1/|\ln N\Lambda L|$ have, of course, been neglected.
The result (\ref {eq:mp}), for $p = 1, \cdots, N{-}1$,
shows that the $N{-}1$ photons
of the ``unbroken'' $U(1)^{N-1}$ gauge group receive non-zero
masses due to nonperturbative effects.%
\footnote
    {
    The vanishing $p=0$ mass corresponds to the extra decoupled photon
    which was added to simplify the duality transformation
    but is not present in the original theory.
    It should be ignored.
    }

\subsection {String tensions}

Let us first examine the vacuum structure of the dual theory in more detail.
The dual scalars are defined to be periodic with period $2 \pi$.
This implies that shifting $\bm\sigma$ by $2\pi$
times any root vector is an identity,
$
    \bm\sigma \equiv \bm\sigma + 2 \pi \alpha_{i}
$
for all $ \alpha_i \in \Delta_{\rm aff}^{0} $.
As noted earlier, the dual action (\ref{eq:Sdual}) is also
invariant under
$
    \bm\sigma \rightarrow \bm\sigma + 2 \pi \mu_i
$,
where $\{ \mu_i \}$ are the fundamental weights of the $SU(N)$ gauge group,
defined by the reciprocity relations (\ref {eq:reciprocity}).
The simple roots $\{ \alpha_i \}$ generate the root lattice $\Lambda_r$.
Its dual, the weight lattice $\Lambda_w$
is generated by the fundamental weights $\{ \mu_i \}$.
The root lattice is a sublattice of the  weight lattice and their quotient is
\begin{equation}
 \Lambda_w/ \Lambda_r= \Z_N \,.
\end{equation}
This implies that the dual theory potential,
$V(\bm\sigma) \equiv -\zeta \, \sum_i \cos(\alpha_i \cdot\bm\sigma)$,
has $N$ isolated minima 
lying within the unit cell of $\Lambda_r$.
These minima are located at 
$\bm\sigma = 0$ and
\begin{equation}
    \bm\sigma = 2\pi\mu_j \,,
    \qquad j = 1, \cdots, N{-}1 \,.
\end{equation}
(Equivalently,
one may describe the minima as lying at
$2\pi \, j \, \mu_1$, for $j = 0, 1, \cdots, N$,
since $\mu_j = j \, \mu_1 + \alpha$ for some $\alpha \in \Lambda_r$.)

Let $\mathcal R$ be some chosen irreducible representation of $SU(N)$.
The expectation value of the Wilson loop $W_{\mathcal R}(C)$ 
characterizes the response of the system to external test charges
in the representation $\mathcal R$.
In a confining phase with a non-zero mass gap,
if external charges in representation $\mathcal R$
cannot be screened by gluons,
then expectation values of large Wilson loops in this representation
are expected to decrease exponentially with the area of the minimal
spanning surface,
\begin{equation}
    \big\langle W_{{\cal R}} (C)\big\rangle
    \sim
    e^{-T({\mathcal R}) \> {\rm Area}(\Sigma) } \,.
\end{equation}
Here $\Sigma$ denotes the minimal surface with boundary $C$,
and $T({\mathcal R})$ is the string tension for representation $\mathcal R$.
Such area law behavior implies the presence of an asymptotically
linear confining potential
between static charges in representation $\mathcal R$ and anti-charges
in representation~$\overline {\mathcal R}$,
$V_{\mathcal R}(\x) \sim T({\mathcal R}) \, |\x|$ as $|\x| \to \infty$.

The irreducible representation $\mathcal R$ may be associated with
its highest weight vector $w \in \Lambda_w$.
Identifying weight vectors which differ by elements of the root lattice
produces a $\Z_N$ grading of representations which corresponds to
their $N$-ality (the charge of the representation under the $\Z_N$ center).
In particular, if this equivalence associates the highest weight vector $w$
with $k$ times the fundamental weight $\mu_1$,
\begin{equation}
    w = k \, \mu_1 + \alpha \,,\qquad \hbox {for some } \alpha \in \Lambda_r \,,
\end{equation}
then the representation $\mathcal R$ has $N$-ality $k$.%
\footnote
    {
    Representations contained in the product of $m$ powers of the
    fundamental representation with $n$ powers of the antifundamental
    have $N$-ality $m{-}n$.
    }

As discussed in Refs.~\cite{Deligne:1999qp,Polyakov:1976fu},
the insertion of a Wilson loop $W_{{\cal R}}(C)$
in a representation $\mathcal R$ with non-zero $N$-ality $k$
corresponds, in the low-energy dual theory,
to the requirement that the dual scalar fields have non-trivial monodromy,
\begin{equation}
    \int_{C'} d \bm\sigma = 2 \pi \mu_k \,,
\label{eq:monodromy}
\end{equation}
where $C'$ is any closed curve whose linking number with $C$ is one.
In other words, in the presence of the Wilson loop $W_{{\cal R}}(C)$
the dual scalar fields must have a discontinuity of $2\pi \mu_k$
across some surface $\Sigma$ which spans the loop $C$.
One way to see this is to go back to the duality relation.
For simplicity, consider the case of a large planar
loop lying in the $xy$-plane.
As the size of the loop grows,
the spanning surface $\Sigma$ approaches an infinite flat plane.
In the presence of the Wilson loop,
the Abelian duality relation $F \sim  * d \sigma$ is replaced by
$F \sim  * d \sigma + \mu_k \, \delta(z) \, dx \wedge dy$.
Therefore the dual scalars $\bm\sigma$ must be discontinuous across
$\Sigma$ in order for the field strength $F$ to be continuous.

The fact that dual low energy theory depends on the representation $\mathcal R$
of the Wilson loop only through its $N$-ality $k$ shows that there
are only $N{-}1$ distinct string tensions,
referred to as $k$-string tensions, $\{ T_k \}$.
(Charge conjugation symmetry implies that $T_k = T_{N-k}$.)
The dual theory representation of Wilson loops also shows that
external charges in representations with zero $N$-ality will not be confined.
These are precisely the representations which can be screened by adjoint
representation gluons.

To evaluate a Wilson loop expectation value,
one must minimize the dual action in the space of field configurations
satisfying the monodromy condition (\ref{eq:monodromy}).
To extract the string tension,
\begin{equation}
    T_k \equiv -\lim_{\mathrm {area}(\Sigma)\to\infty}
    \frac{\ln\left\langle W_{\mathcal R}(C)\right\rangle}
    {\mathrm {area}(\Sigma)} \,,
\end{equation}
it is sufficient to consider the limit where $\Sigma$ fills the $xy$-plane.
In this case, the field $\bm \sigma(\x)$ will only depend on $z$.
It must approach some minimum of the dual potential at infinity,
$\lim_{z\to\pm\infty} \bm \sigma(z) = 2\pi\mu_l$,
and must be discontinuous across $z=0$ with a jump given by the
prescribed fundamental weight,
$
    \lim_{z\to 0^+} \bm\sigma(z) - \lim_{z\to 0^-} \bm\sigma(z)
    =
    2 \pi \mu_k \pmod {2\pi} 
$.
Because shifts by $2\pi\mu_k$ are an invariance of the dual potential
$V(\bm\sigma)$, one may equally well minimize the action for field
configurations $\bm\sigma(z)$ which are continuous but whose asymptotic
values differ,
\begin{equation}
    T_k =
    \left.
    \min_{\bm\sigma(z)} \;
    \frac {\Delta S(\sigma)}{\mathrm{area}(\R^2)}
    \right|_{\Delta\bm\sigma = 2\pi\mu_k \;(\mathrm {mod}\; 2\pi)}
    \,,
\end{equation}
where
$
    \Delta\bm\sigma \equiv \bm\sigma(\infty) - \bm\sigma(-\infty)
$,
and $\Delta S(\sigma)$ is the dual action minus its vacuum value.
Explicitly,
\begin{equation}
    T_k
    =
    \left.
    \min_{\bm\sigma(z)} \;
    \int dz \;
    \Bigl\{
	\frac{1}{2L}
	\left(\frac{g}{2\pi}\right)^2
	\left(\frac{\partial \bm \sigma}{\partial z} \right)^2
	+ \zeta \, \sum_i \left[1-\cos (\sigma_i {-} \sigma_{i+1})\right]
    \Bigr\}
    \right|_{\Delta\bm\sigma = 2\pi\mu_k \;(\mathrm {mod}\; 2\pi)}
    \,.
\label{1daction}
\end{equation}
In other words, the $k$-string tension $T_k$ equals the action of a kink
solution with topological charge $k$ in this one dimensional theory.

The width of the kink solution must be of order of the inverse photon mass
$m_\gamma ^{-1}$.
Consequently, the $k$-string tension will have the form
$
    T_k = f_k \> T
$,
where
\begin{equation}
    T \equiv \zeta / m_\gamma 
    \sim
    \Lambda^2 (\Lambda LN)^{-1/6} \left| \log (\Lambda LN)\right|^{-3/11}  
    \,,
\label {eq:Tdef}
\end{equation}
and $f_k$ is an $O(1)$ coefficient.
Even without finding the minimizing kink solutions explicitly,
it is apparent that the resulting $k$-string tension $T_k$ will be
non-zero (for $k = 1, \cdots, N{-}1$),
and must satisfy the convexity relation
$
    T_{k+l} \le T_k + T_l
$.

We were unable to solve the kink equations of motions analytically for
general $N$,
but when $N=2$ the equations of motion reduce to Sine-Gordon model.
In this case, one finds
\begin{equation}
    T \equiv T_1 = 4 \sqrt 2  \, \zeta/m_\gamma \,.
\end{equation}

\subsection{Larger size or larger $N$}

The above semiclassical analysis of the deformed Yang-Mills theory
is reliable provided there is a parametrically large separation of scales
between the lightest $W$-boson mass, $m_W = 2\pi/(NL)$,
and the nonperturbatively induced dual photon mass $m_\gamma$.
Their ratio scales as
\begin{equation}
    \frac{m_{W}}{m_\gamma}
   \sim (LN \Lambda)^{-11/6} \> | \log (L N \Lambda)|^{-9/11}   \,,
\end{equation}
and hence there is a large separation of mass scales provided $LN\Lambda \ll 1$.
In this regime,
the monopole gas is highly dilute and a semiclassical analysis
is justified.
Increasing $LN\Lambda$,  by increasing $N$, $L$, or both,
decreases the separation of scales;
the heaviest photon mass, $m_\gamma$,
grows while the lightest $W$ mass, $m_W$, drops.
When $LN\Lambda \approx 1$, the scale separation is entirely lost,
the effective 't Hooft coupling $\lambda \equiv g^2 N$ at the scale
of $m_W$ ceases to be small,
and the long distance dynamics can no longer be described by
a weakly coupled $U(1)^{N-1}$ effective theory.

One can consider sending $N$ to infinity while staying within
the analytically tractable regime.
This is a double scaling limit in which
$g^2 N$ and $L N \Lambda$ are both are held fixed
(and both are much less than unity)
as $N \to\infty$.
Taking a large $N$ limit in this fashion allows
monopole effects to survive and to continue dictating
the nonperturbative physics of the deformed Yang-Mills theory.
However, this region shrinks to a vanishingly small window in the
large $N$ limit,
since the double scaling implies that
$0 < L \ll L_{\rm max}$ with
$L_{\rm max} \, \Lambda \sim  1/N$.
For any fixed compactification size $L$,
if one sends $N\to\infty$ the deformed YM theory
ceases to possess a monopole dominated,
Abelian long distance regime.%
\footnote
    {
    An analog of this double scaled limit was previously discussed
    by Douglas and Shenker in mass
    deformed $SU(N)$ $\N=2$ supersymmetric Yang-Mills theory on $\R^4$
    \cite{Douglas:1995nw}.
    This deformed $\N=1$ supersymmetric theory,
    just like our deformed Yang-Mills theory,
    possess a regime  in which the long distance gauge dynamics reduces
    to the Abelian subgroup $U(1)^{N-1}$.
    Ref.~\cite{Douglas:1995nw} shows that in the $N \rightarrow  \infty $
    limit of the mass deformed theory,
    the  Abelian long distance regime is preserved only if the
    mass deformation $m$ is sent to zero in a correlated fashion,
    $m/\Lambda \sim {1}/{N^4} $.
    In particular, at any fixed non-zero $m$, if one takes $N \to \infty$
    first then there is no regime of the supersymmetric gauge theory in
    which the long distance dynamics remains Abelian.
    Although this phenomena only appeared previously in the context
    of supersymmetric gauge theories, it is generic in deformed Yang-Mills
    and other deformed QCD-like theories.
    }

\subsection{Connection to integrable Toda theory} 
\label{sec:toda}

The $\Z_N$ symmetric model (\ref{eq:Sdual})
is a deformation of a complex affine Toda theory with action
\begin{equation}
    S^{\rm affine \; Toda}
    =
    \int_{\R^3} \;
    \Big[
	\frac{1}{2L}
	\left(\frac{g}{2\pi}\right)^2
	(\nabla \bm\sigma)^2
	-
	\zeta \>
	\sum_{i=1}^N \> e^{ i(\sigma_i- \sigma_{i+1})}
    \Big] \,.
\end{equation}
This complex (CPT-noninvariant) action describes a plasma
which is composed solely of monopoles with no antimonopoles.
(Due to the existence of the affine root, one can
have a neutral plasma composed solely of monopoles!)
Interestingly, the soliton spectrum of the affine Toda theory
is exactly computable.
When reduced to one dimension, this theory
is an integrable system as shown
by Hollowood \cite{Hollowood:1992by}, using techniques
due to Hirota \cite{hirota, toda}. 

As discussed above,
the $k$-string tension $T_k$ is equal 
to the action of the kink solution with topological charge $k$.
Borrowing the exact soliton spectrum from Ref.~\cite{Hollowood:1992by},
one finds that the $k$-string tensions in the affine Toda theory are given by
\begin{equation}
    T_k^{\rm affine \; Toda} =  T  N \, \sin \frac{\pi k}{N} \,,
    \qquad  k=1, N{-}1 \,,
\label{sinelaw}
\end{equation}
with $T$ given above in Eq.~(\ref{eq:Tdef}).  

The long distance effective theory (\ref{eq:Sdual})
for our deformed Yang-Mills theory
(when $LN\Lambda \ll 1$)
is a deformation of the affine Toda system by complex conjugation.
Unfortunately, unlike the integrable affine Toda system,
when $N > 2$ the resulting CPT invariant system
is no longer exactly integrable according to Hirota's criteria.%
\footnote
    {
    In the absence of the complex conjugate term in the potential,
    there is a  change of variables which converts the soliton equation
    of motion into ``Hirota bilinear type,"
    which is synonymous with solvability \cite{Hollowood:1992by}.   
    The presence of the complex conjugate term spoils the bi-linearity. 
    }
Consequently, we do not expect $k$-string tensions
in the deformed Yang-Mills theory to have the sine-law form (\ref{sinelaw}).

Recently, there have been attempts \cite{Diakonov:2007nv},
to model the {\it strongly}
coupled confined regime of Yang-Mills theory
assuming the Wilson line has the center-symmetric form (\ref{eq:vev}).
(See also earlier related work in
Refs.~ \cite{Lee:1997vp, Kraan:1998pm, Bruckmann:2004nu, Davies:2000nw}.)
A few remarks concerning the connection with Ref.~\cite{Diakonov:2007nv}
may be in order.
First, our results for the $k$-string tensions do not support
the claim of Ref.~\cite{Diakonov:2007nv}, which asserts that
$k$-string tensions will have the sine law form (\ref{sinelaw}).
As just noted, sine-law string tensions are a property of
the affine Toda subsystem, whereas the center-stabilized Yang-Mills in
a weak coupling regime is dual to a real deformation
of the affine Toda theory.
We see no reason to believe that the $k$-dependence of the string tensions
will be unaffected by the deformation.
Secondly, it should be emphasized that
the deformation (\ref{eq:P[Omega]}) stabilizes the center symmetric vacuum 
in the {\it weakly} coupled regime, and thereby provides a window in which 
a semiclassical analysis is reliable.
Many earlier discussions of center symmetric backgrounds do not
clearly distinguish the weakly coupled ``Higgs'' regime, in which 
fluctuations of the Wilson line eigenvalues are small,
from the strong coupling regime in which the eigenvalues
have large fluctuations and are essentially randomized
over the unit circle.
In our deformed Yang-Mills theory, both regimes exist.
As the compactification size $L$ increases, the theory
moves from the weakly coupled regime to the strongly coupled regime.
These two regimes are expected to be smoothly connected ---
no physical order parameter sharply distinguishes the two regimes.
Nevertheless, the long distance physics of the weak coupling Higgs regime
is effectively Abelian, while in the strong coupling regime there is
no length scale beyond which the dynamics can be described accurately
in terms of Abelian degrees of freedom.

\acknowledgments
We thank Barak Bringoltz and Steve Sharpe for calling our attention
to the issue of spontaneous breaking to diagonal subgroups
when there are multiple compactified directions.
This work was supported in part by the U.S.~Department of Energy
under Grant Nos.~DE-AC02-76SF00515 and DE-FG02-96ER40956,
and by the U.S.~National Science Foundation under Grant No.~PHY05-51164.
L.Y. thanks the Kavli Institute for Theoretical Physics for its
hospitality during the completion of this paper.

\newpage
\appendix

\section{Monopole measure}
\label{sec:B}

The appropriate one-loop measure for integrating over configurations of a
single monopole (of any type) may be expressed as%
\footnote
    {
    The following summary is an adaptation of the appendix of
    Ref.~\cite{Davies:2000nw}, which treats the monopole measure
    in supersymmetric Yang-Mills theory.
    }
\begin{eqnarray}
    d \mu_{\rm monopole}
    =
    \mu^{4} \, e^{-(S_{\rm YM} + \Delta S)} \,
    \frac{d^3 \bm a}{(2 \pi)^{3/2}}  \; J_{\bm a} \;
    \frac{d \phi}{(2 \pi)^{1/2}} \; J_{\phi} \;
    \left[{\rm det'}(-D^2_{\rm adj}) \right]^{-1}  \,,
\end{eqnarray}
where $\bm a \in \R^3$ is the monopole position,
$\phi \in [-\pi,\pi]$ is the internal $U(1)$ angle of the monopole,
and $\mu$ is the (Pauli-Villars) renormalization scale.
Global $U(1)$ gauge transformations
(in the $U(1)$ subgroup associated with the given type of monopole)
shift the angle $\phi$.
Fluctuations in the position and $U(1)$ angle of the monopole
represent the four zero modes in the monopole small fluctuation operator;%
\footnote
    {
    For comparison, recall that an instanton in $SU(N)$ Yang-Mills theory
    has $4N$ zero-modes.
    (See, for example, Ref.~\cite{Weinberg:2006rq}.)
    For $SU(2)$ gauge theory, these are the four zero modes corresponding
    to changes in the instanton position,
    one for its size,
    and three global gauge rotations.
    For $SU(N)$ with $N > 2$, there are in addition $4N{-}8$ zero modes
    associated with changes in the embedding of $SU(2)$ within $SU(N)$.
    When compactified in one dimension, with non-trivial holonomy $\Omega$,
    one may regard an instanton as being composed of $N$ independent
    monopole constituents \cite{Kraan:1998pm, Lee:1997vp},
    each of which carries four zero-modes.
    }
the factor of $\mu^{4}$ can be viewed as the contributions of the
Pauli-Villars regulator fields associated with these bosonic zero modes.
The exponential factor is, of course, the exponential of minus the
classical action of the monopole.
The collective coordinate Jacobians are given by \cite{Davies:2000nw}  
\begin{equation}
    J_a = S^{3/2}_{\rm YM} \,, \qquad
    J_{\Omega} =
    \frac{2\pi \, S^{1/2}_{\rm YM}}{\alpha_j \cdot \lambda}
    =  N  L \, S^{1/2}_{\rm YM}  \,,
\end{equation}
where $\lambda = \{ 0, 2\pi, 4\pi, \cdots, 2\pi(N{-}1) \}/(NL)$
are the eigenvalues of $-i\ln\Omega$.
The primed determinant represents the result of Gaussian
integrals over all fluctuations other than zero modes;
the prime on the determinant denotes omission of the zero modes.
The contributions from gauge bosons and ghosts,
\begin{equation}
    {\underbrace{{\left[ {\det}'
	( -D^2 \delta_{\mu  \nu} - 2 F_{\mu \nu})_{\rm adj}
    \right ]^{-1/2} }}_{\rm gauge \;\; bosons}}
    \times
    \underbrace{ {\det} (D^2)_{\rm adj} }_{\rm ghosts} \,,
\end{equation}
combine to give this simple form because
\begin{equation}
    [{\det}^{'} ( -D^2 \delta_{\mu  \nu} - 2 F_{\mu \nu})_{\rm adj}]^{-1/2}
    = [{\det} ( -D^2)_{\rm adj}]^{-2}
\end{equation}
in any self-dual background.
These functional determinants may be regularized using the Pauli-Villars scheme.

The fields of fundamental monopoles reside entirely within an $SU(2)$
subgroup of $SU(N)$, and
the characteristic size of these monopoles is given by the inverse
of the lightest $W$-boson mass, $m_W^{-1} \sim NL$.
(This is the only scale which appears in the classical equations
for the monopole.)
The regularized scalar determinant depends on the cube root of
the renormalization scale,
$\det(-D^2) \sim \mu^{1/3}$.
Since the determinant is dimensionless, it must have the form
\begin{equation}
    [\det(-D^2)]^{-1} = 2\pi C \, (\mu NL)^{-1/3} \,,
\end{equation}
where $C$ is a pure number ($N$-independent).
Consequently,
the one-loop monopole measure equals
\begin{equation}
    d \mu_{\rm monopole}
    =
    C
    \mu^{11/3} \,
    (N L)^{\frac{2}{3}} \,
    (S_{\rm YM})^2 \,
    e^{-{S_{\rm YM} + \Delta S}} \;
    d^3 {\bm a} \; d\Omega \,.
\label{Eq:measure}
\end{equation}
Performing the trivial integral over the angle $\Omega$,
the result is the $\zeta \; d^3 {\bm a}$,
with $\zeta$ the monopole fugacity.
Choosing to use $m_W$ as the value of the renormalization point
yields the expression (\ref{eq:fug}) for the fugacity.

\sloppy

\end{document}